\documentclass[iop]{emulateapj}

\usepackage{ulem}

\newcommand\kms{\ifmmode{\rm km\thinspace s^{-1}}\else km\thinspace s$^{-1}$\fi}
\newcommand\gaia{{\it Gaia\/}}
\newcommand\hip{{\it Hipparcos\/}}
\newcommand\hstar{GJ\,67}
\newcommand\jaavso{Journal of the American Association of Variable Star Observers}

\shortauthors{Torres}
\shorttitle{\hstar}

\begin{document} 
\submitted{Accepted for publication in Monthly Notices of the Royal Astronomical Society}

\title{Orbital solution and dynamical masses for the nearby binary system \hstar\ AB}

\author{
Guillermo Torres
}

\affiliation{Center for Astrophysics $\vert$ Harvard \&
  Smithsonian, 60 Garden St., Cambridge, MA 02138, USA;
  gtorres@cfa.harvard.edu}

\begin{abstract}
We report spectroscopic observations of the nearby, 19.5\,yr
binary system \hstar AB spanning more than 35 yr. We carry out a
global orbital solution combining our
radial velocity measurements with others from the literature
going back more than a century, and with all other available
astrometric observations. The latter include measurements of the
relative position as well as the \hip\ intermediate
data and photographic observations tracing the motion of
the photocentre. We derive masses for the primary and
the M dwarf secondary of $0.95 \pm 0.11$ and
$0.254 \pm 0.019$\,M$_{\sun}$, respectively, as well as a more accurate
trigonometric parallax of $79.08 \pm 0.63$~mas that accounts for
the orbital motion. We provide evidence suggesting that the much smaller
parallax from \gaia\ DR3 is biased. The precision in the masses remains limited mainly
by the still few measurements of the relative position.
\end{abstract}


\section{Introduction}
\label{sec:introduction}

\hstar\ (WDS J01418+4237AB, HD~10307, HIP 7918, HR 483) is a 5th magnitude star only $\sim$13~pc away, with
properties are very similar to the Sun. The spectral classification
has typically been given as \ion{G1.5}{5} or \ion{G2}{5}. Its binary
nature was discovered in a proper motion study based on photographic
measurements with 61-inch refactor at the Sproul Observatory, and was
first announced by \cite{Lippincott:1976}. In a subsequent
investigation \cite{Lippincott:1983} reported an astrometric orbital
solution for the motion of the centre of light, with a period of
19.5~yr and a semimajor axis of 0\farcs13. The same study also
resolved the pair spatially for the first time in the near-infrared by
the technique of speckle interferometry. The companion is a mid-to-late M dwarf.

Estimates of the mass of the components have typically relied on
various assumptions or external information. \cite{Lippincott:1983}
reported values of $1.44 \pm 0.35$ and $0.38 \pm
0.07$\,M$_{\sun}$ for the primary and secondary, whereas
\cite{Henry:1993} gave $0.93 \pm 0.23$ and $0.280 \pm
0.071$\,M$_{\sun}$, respectively. \cite{Martin:1998} found much lower
values of $0.80 \pm 0.16$ and $0.136 \pm 0.053$\,M$_{\sun}$.

\hstar\ has a long history of radial velocity measurements dating back
more than a century. Spectroscopic orbital elements have been reported
by \cite{Duquennoy:1991} and \cite{Abt:2006}, but relied in part on
elements adopted from the astrometry of \cite{Lippincott:1983}.  The
first spectroscopic orbit based solely on radial velocities is more
recent \citep{Fekel:2018}, and used those authors' own measurements
combined with others from the literature to derive much improved
elements.

In a separate effort we have been monitoring the radial velocities of
\hstar\ at the Center for Astrophysics for more than 35~yr, spanning
close to two orbital cycles of the binary. Additionally, a handful of
astrometric observations that resolve the companion have
appeared since the study of \cite{Lippincott:1983}, and were used by
\cite{Miles:2017} to infer preliminary elements for the relative
orbit. The individual Lippincott observations themselves have never
been used in any other orbital analysis beyond the original study.
\hstar\ was also observed by the \hip\ mission \citep{ESA:1997},
and while the observations did not resolve the binary, the
intermediate astrometric measurements for the star are available to
strengthen the determination of the photocentre orbit. The existence
of all of this observational material, plus a further series of
radial-velocity measurements published after the paper by
\cite{Fekel:2018}, motivates us to carry out the first global orbital
solution that combines all available observations in a self-consistent
manner.  

Section~\ref{sec:spectroscopy} reports our spectroscopic observations
of \hstar\ using three different instruments.
Section~\ref{sec:astrometry} summarizes the existing astrometric
observations.  Our orbital analysis is presented in
Section~\ref{sec:analysis}, followed by a discussion of results in
Section~\ref{sec:discussion}, and conclusions.

\section{Radial velocity measurements}
\label{sec:spectroscopy}

Our spectroscopic observations of \hstar\ at the Center for
Astrophysics (CfA) began in September of 1986, and used an echelle
spectrograph \citep[Digital Speedometer (DS);][]{Latham:1992} on the
1.5m Wyeth reflector at the (now closed) Oak Ridge Observatory
(Massachusetts, USA). This instrument had a resolving power of $R
\approx 35,\!000$, and was equipped with an intensified photon-counting
Reticon detector that limited the recorded output to a single order
45\,\AA\ wide centred at 5187\,\AA, featuring the \ion{Mg}{1}~b
triplet. Typical signal-to-noise ratios were about 45 per resolution
element of 8.5\,\kms, and a total of 53 spectra were obtained regularly
through September of 2004. Two additional observations were gathered
near the end of 2009 with a nearly identical instrument attached to
the 1.5m Tillinghast reflector at the Fred L.\ Whipple Observatory
(Arizona, USA).

Starting in December of 2009 the observations were continued on the Tillinghast
reflector with a modern, bench-mounted, fibre-fed instrument
\citep[Tillinghast Reflector Echelle Spectrograph
  (TRES);][]{Szentgyorgyi:2007, Furesz:2008} providing a resolving
power of $R \approx 44,\!000$. These spectra cover the wavelength range
3800--9100\,\AA\ in 51 orders. For the order centred at about
5187\,\AA\ the typical signal-to-noise ratios of the 48 observations we
collected through November of 2021 were about 200 per resolution element
of 6.8\,\kms.

Wavelength solutions were based on exposures of a thorium-argon lamp
before and after each science exposure. For the DS observations, the
velocity zero point was monitored by means of sky exposures at dusk
and dawn, and small run-to-run corrections usually smaller than
2~\kms\ were applied to the raw velocities. Observations of minor
planets were then used to determine a further correction of
+0.14\,\kms\ to the IAU system \citep[see][]{Stefanik:1999}. For TRES, the
much smaller drifts in the velocity zero point ($\leq 100$~m~s$^{-1}$)
were monitored by observing IAU standards each run, and asteroid
observations were again used to transfer the velocities to the IAU
system.

The binary companion of \hstar\ is very faint, and all our spectra are
therefore single-lined.  Radial velocities were measured by
cross-correlation using the {\tt XCSAO} task running under
{\tt IRAF}.\footnote{{\tt IRAF} is distributed by the National Optical Astronomy
  Observatory, which is operated by the Association of Universities
  for Research in Astronomy (AURA) under a cooperative agreement with
  the National Science Foundation.} The template was taken from a
large library of synthetic spectra based on model atmospheres by
R.\ L.\ Kurucz, and a line list tuned to better match real stars
\citep[see][]{Nordstrom:1994, Latham:2002}. To determine the best
parameters for the template, we first used our higher-quality TRES
spectra to estimate the spectroscopic parameters employing the SPC
procedure \citep[Stellar Parameter Classification;][]{Buchhave:2012}.
This procedure compares the observed spectra against the spectral
library, and for each observation it selects the spectroscopic parameters
giving the highest cross-correlation coefficient from a
multi-dimensional fit. The four parameters are the effective temperature
$T_{\rm eff}$, the surface gravity $\log
g$, the metallicity [m/H], and the rotational broadening $v \sin i$. We
averaged the spectroscopic properties over all spectra, and obtained $T_{\rm eff} =
5854 \pm 50$\,K, $\log g = 4.31 \pm 0.10$, ${\rm [m/H]} = -0.04 \pm
0.08$, and $v \sin i = 3.0 \pm 1.0$\,\kms. These properties are fairly
consistent with other independent determinations in the literature
\citep[see, e.g.,][]{Allende:2004, Ramirez:2007, Boeche:2016}. For the
radial-velocity measurements we chose a template with parameters in
our library near these values: $T_{\rm eff} = 6000$~K, $\log g = 4.5$,
solar [m/H], and $v \sin i = 2$\,\kms. For consistency with the DS
spectra, we restricted the cross-correlations to the TRES order
containing the Mg triplet.

\setlength{\tabcolsep}{6pt}  
\begin{deluxetable*}{lccc}[!t]
\tablewidth{0pc}
\tablecaption{Radial Velocities for \hstar\ from the CfA Digital Speedometers\label{tab:ds}}
\tablehead{
\colhead{HJD} &
\colhead{$RV$} &
\colhead{$\sigma_{\rm RV}$} &
\colhead{Orbital Phase}
\\
\colhead{(2,400,000+)} &
\colhead{(\kms)} &
\colhead{(\kms)} &
\colhead{}
}
\startdata
 46685.7187  &  4.72  &  0.36  &  0.4645 \\
 47813.6325  &  4.63  &  0.36  &  0.6225 \\
 48170.7243  &  3.92  &  0.42  &  0.6726 \\
 48558.7617  &  2.80  &  0.65  &  0.7269 \\
 48602.4871  &  2.98  &  0.60  &  0.7330
\enddata
\tablecomments{Orbital phases were computed from the ephemeris given in
  Section~\ref{sec:analysis}. (This table is available in its entirety
  in machine-readable form.)}
\end{deluxetable*}
\setlength{\tabcolsep}{6pt}  

\setlength{\tabcolsep}{6pt}  
\begin{deluxetable*}{lccc}[!t]
\tablewidth{0pc}
\tablecaption{CfA Radial Velocities for \hstar\ from TRES\label{tab:tres}}
\tablehead{
\colhead{HJD} &
\colhead{$RV$} &
\colhead{$\sigma_{\rm RV}$} &
\colhead{Orbital Phase}
\\
\colhead{(2,400,000+)} &
\colhead{(\kms)} &
\colhead{(\kms)} &
\colhead{}
}
\startdata
 55166.6932  &  3.98  &  0.04  &  0.6527 \\
 55193.5961  &  3.99  &  0.05  &  0.6565 \\
 55549.6456  &  3.57  &  0.03  &  0.7064 \\
 55585.5598  &  3.52  &  0.03  &  0.7114 \\
 55884.7453  &  3.08  &  0.04  &  0.7533
\enddata
\tablecomments{Orbital phases were computed from the ephemeris given in
  Section~\ref{sec:analysis}. (This table is available in its entirety
  in machine-readable form.)}
\end{deluxetable*}
\setlength{\tabcolsep}{6pt}  

The radial velocities for the DS and TRES may be found in
Tables~\ref{tab:ds} and \ref{tab:tres}, respectively, along with
internal error estimates produced by {\tt XCSAO}.

\subsection{Radial Velocities from the Literature}
\label{sec:literature_rvs}

\begin{figure*}
\epsscale{1.18}
\plotone{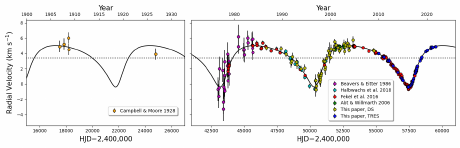}
\figcaption{Radial-velocity measurements for \hstar\ from various
  sources, as labelled. The solid curve is our best-fitted model described
  in Section~\ref{sec:analysis}, and the dotted line represents the
  centre-of-mass velocity for the system. \label{fig:rvs}}
\end{figure*}

In addition to our own 55 velocities of \hstar\ from the DS and 48
from TRES, the analysis of Section~\ref{sec:analysis} incorporates five
much older velocities from the Lick Observatory \citep{Campbell:1928},
32 from the E.\ W.\ Fick Observatory \citep{Beavers:1986}, 17 measurements by \cite{Abt:2006} from the Kitt Peak Observatory, 157 velocities published by
\cite{Fekel:2018} obtained with several instruments at the Kitt Peak and
Fairborn Observatories, and 32 velocities from the
CORAVEL spectrometer at the Observatory of Haute-Provence
\citep{Halbwachs:2018}, which were not available to \cite{Fekel:2018} for
their own spectroscopic study.
All of these velocities are shown graphically
in Fig.~\ref{fig:rvs}, along with our best-fitted model described
later. The total time span of the measurements is 115~yr (about 5.9
orbital cycles).

Formal uncertainties for the velocities were taken as published, for
the sources that reported them. For the Lick velocities we adopted
errors of 0.5\,\kms, and for the Kitt Peak measurements of \cite{Abt:2006} we used
0.21~\kms, following those authors. The observations by
\cite{Fekel:2018} were reported with weights rather than uncertainties.
We converted the weights to uncertainties by adopting 0.1\,\kms\ as the
error for an observation of unit weight.
As formal errors can sometimes be underestimated or
overestimated, all radial-velocity uncertainties were adjusted during
our analysis as described below.

\section{Astrometric Observations}
\label{sec:astrometry}

\subsection{Photocentre Motion from Photographic Plates}

The photographic observations of \hstar AB from the Sproul Observatory
that formed the basis of the study by \cite{Lippincott:1983} were
taken between 1937 and 1982. While the original plate measurements were
never published, those authors did report normal point residuals from
their solution for parallax, proper motion, and orbital motion at 18
epochs, given to a precision corresponding to about 2~mas. Based on
the information provided, it is possible to reconstruct observations
that reflect the motion of the centre of light of the binary around
the barycentre in the right ascension and declination directions
($\Delta X$, $\Delta Y$). Given the faintness
of the companion at the wavelength of the photographic observations,
for all practical purposes the photocentre coincides with the primary.
We used these reconstructed observations as measurements for our
analysis.  Formal uncertainties $\sigma_{\Delta X}$ and
$\sigma_{\Delta Y}$ were calculated from the reported
standard error of unit weight in each coordinate, and the total weight
assigned to each observation by \cite{Lippincott:1983}. These
reconstructed $\Delta X$ and $\Delta Y$ observations are given in
Table~\ref{tab:lippincott}.

\setlength{\tabcolsep}{6pt}  
\begin{deluxetable*}{lccccc}
\tablewidth{0pc}
\tablecaption{Reconstructed Photocentre Observations for \hstar AB from
  \cite{Lippincott:1983}\label{tab:lippincott}}
\tablehead{
\colhead{Date} &
\colhead{$\Delta X$} &
\colhead{$\sigma_{\Delta X}$} &
\colhead{$\Delta Y$} &
\colhead{$\sigma_{\Delta Y}$} &
\colhead{Orbital Phase}
\\
\colhead{(year)} &
\colhead{(\arcsec)} &
\colhead{(\arcsec)} &
\colhead{(\arcsec)} &
\colhead{(\arcsec)} &
\colhead{}
}
\startdata
 1938.48  &  $-$0.042\phs  &  0.015  &  $-$0.064\phs  &  0.014  &  0.9972 \\
 1940.35  &     0.023      &  0.012  &  $-$0.014\phs  &  0.011  &  0.0929 \\
 1941.66  &     0.077      &  0.015  &     0.037      &  0.014  &  0.1599 \\
 1963.52  &     0.072      &  0.010  &     0.135      &  0.009  &  0.2785 \\
 1964.83  &     0.094      &  0.013  &     0.149      &  0.013  &  0.3456 \\
 1965.82  &     0.106      &  0.015  &     0.150      &  0.014  &  0.3962 \\
 1968.80  &     0.080      &  0.007  &     0.142      &  0.007  &  0.5487 \\
 1969.88  &     0.049      &  0.006  &     0.131      &  0.005  &  0.6040 \\
 1970.80  &     0.038      &  0.007  &     0.131      &  0.007  &  0.6511 \\
 1973.81  &  $-$0.007\phs  &  0.007  &     0.031      &  0.007  &  0.8051 \\
 1974.83  &  $-$0.031\phs  &  0.008  &  $-$0.019\phs  &  0.008  &  0.8573 \\
 1975.75  &  $-$0.048\phs  &  0.008  &  $-$0.021\phs  &  0.008  &  0.9044 \\
 1976.76  &  $-$0.058\phs  &  0.006  &  $-$0.059\phs  &  0.005  &  0.9561 \\
 1977.94  &  $-$0.016\phs  &  0.008  &  $-$0.056\phs  &  0.007  &  0.0164 \\
 1978.80  &     0.012      &  0.006  &  $-$0.031\phs  &  0.005  &  0.0604 \\
 1979.77  &     0.033      &  0.007  &     0.002      &  0.006  &  0.1101 \\
 1980.74  &     0.052      &  0.006  &     0.051      &  0.006  &  0.1597 \\
 1982.09  &     0.088      &  0.007  &     0.092      &  0.006  &  0.2288 
\enddata
\tablecomments{Orbital phases were computed from the ephemeris given in
  Section~\ref{sec:analysis}.}
\end{deluxetable*}
\setlength{\tabcolsep}{6pt}  

\subsection{Relative Positions}
\label{sec:relative}

Only a few successful measurements of the relative position of the
components in \hstar AB have been made in the 40 years since the first
of them by \cite{Lippincott:1983}, mostly at near-infrared
wavelengths. About as many attempts have failed to resolve the pair,
either because of the faintness of the companion at the chosen
wavelengths or because the separation was too small. A listing of all
of these measurements, contained in the Washington Double Star Catalogue
\citep[WDS;][]{Worley:1997, Mason:2001}, was kindly provided by
R.\ Matson (US Naval Observatory) and is reproduced in
Table~\ref{tab:wds}. We include the bandpass and magnitude
difference $\Delta m$ measured for each observation.

\setlength{\tabcolsep}{6pt}  
\begin{deluxetable*}{lcccccl}
\tablewidth{0pc}
\tablecaption{Relative Positions of \hstar AB from the WDS \label{tab:wds}}
\tablehead{
\colhead{Date} &
\colhead{P.A.} &
\colhead{Separation} &
\colhead{$\lambda$} &
\colhead{$\Delta m$} &
\colhead{Orbital Phase} &
\colhead{Reference}
\\
\colhead{(year)} &
\colhead{(deg)} &
\colhead{(\arcsec)} &
\colhead{($\mu m$)} &
\colhead{(mag)} &
\colhead{} &
\colhead{}
}
\startdata
 1982.755   &       N--S                &  $0.550 \pm 0.028$          & $K$ & $3.6 \pm 0.2$   &  0.2628  &  \cite{Lippincott:1983} \\
 1982.755   &       E--W                &  \phm{:}$0.38 \pm 0.04$:    & $K$ & $3.6 \pm 0.2$   &  0.2628  &  \cite{Lippincott:1983} \\
 1989.773   &  $203 \pm 2$\phn\phn      &  $0.623 \pm 0.027$          & $K$ & $4.30 \pm 0.07$ &  0.6220  &  \cite{Henry:1993} \\
 1990.906   &  $199 \pm 2$\phn\phn      &  $0.442 \pm 0.018$          & $K$ & $4.50 \pm 0.12$ &  0.6799  &  \cite{Henry:1993} \\
 1990.915   &  $194 \pm 2$\phn\phn      &  $0.448 \pm 0.030$          & $J$ & $4.37 \pm 0.25$ &  0.6804  &  \cite{Henry:1993} \\
 1990.931   &  $196 \pm 2$\phn\phn      &  $0.485 \pm 0.025$          & $K$ & $4.50 \pm 0.05$ &  0.6812  &  \cite{Henry:1993} \\
 1995.7587  &  ($190.3:$)               &  \phm{:}$0.306 \pm 0.020$:  & $V$ & \nodata         &  0.9283  &  \cite{Hartkopf:1997} \\
 2002.7760  &  $212.0 \pm 1.0$\phn\phn  &  $0.75 \pm 0.02$            & $I$ & $5.5 \pm 0.4$   &  0.2874  &  \cite{Roberts:2011} \\
 2006.687   &  \nodata                  &  $0.78 \pm 0.04$            & $K_{\rm S}$ & 3.9     &  0.4875  &  \cite{Serabyn:2007}
\enddata
\tablecomments{Orbital phases were computed from the ephemeris given in
  Section~\ref{sec:analysis}. The 1995 P.A.\ in parentheses was not used.
  The HD identifier given in the original paper for the 2002 observation (HD 10105) is incorrect.}
\end{deluxetable*}
\setlength{\tabcolsep}{6pt}  

A few notes about these measurements are in order. The 1982 speckle
observation by \cite{Lippincott:1983} was reported split into a
north-south separation with an associated uncertainty, and a more
uncertain east-west separation without an error. We have accordingly
assigned a more conservative uncertainty to the latter. The 1995 speckle
observation by \cite{Hartkopf:1997} was given without errors, and
considered uncertain. A preliminary analysis revealed that the
position angle is indeed in error by more than 30\arcdeg, and we have
therefore discarded it. We assigned a reasonable uncertainty to the
angular separation.  Typical uncertainties were also adopted for the
2002 adaptive optics observation by \cite{Roberts:2011}. Finally, the
2006 adaptive optics observation of \cite{Serabyn:2007} only yielded
an angular separation.  Even though the 1982, 1995, and 2006
observations are incomplete in the sense of not providing a position
angle paired with a separation, they still contain valuable
information in this case because of the very small overall number of
measurements. We have therefore made use of them in our analysis.

The most recent solution for the astrometric orbit of \hstar AB was
published by \cite{Miles:2017}. That study used all six complete
observations in Table~\ref{tab:wds} (with both a position angle and a
separation), plus one additional measurement that turns out to be
spurious, and unfortunately biased their results. That observation has
since been removed from the WDS, and we do not list it in our table.

\subsection{\hip\ Intermediate Data}

The \hip\ mission observed \hstar AB (HIP~7918) a total of 52 times
between December of 1989 and January of 1993, but did not resolve the
companion. Therefore the measurements refer to the centre of light of
the system.  The five-parameter solution to derive the position
($\alpha_0^*$, $\delta_0$), proper motion ($\mu_{\alpha}^*$,
$\mu_{\delta}$)\footnote{Following the practice in the \hip\
  catalogue we define $\alpha^* \equiv \alpha \cos\delta$ and
  $\mu_{\alpha}^* \equiv \mu_{\alpha} \cos\delta$.}, and trigonometric
parallax ($\pi_{\rm t}$) of the object as reported in the original
catalogue \citep{ESA:1997} did not take into account the orbital motion of the
binary, and as a result some or all of those parameters may be
affected.  The parallax is of particular interest for this work
because it factors into the determination of the total mass of the
system. The re-reduction of the satellite data carried out a decade
later by \cite{vanLeeuwen:2007} also did not account for orbital
motion.

The intermediate astrometric data from the mission are publicly
available in the form of ``abscissa residuals'', i.e., residuals from
the five-parameter solution reported in the catalogue. These
one-dimensional measurements, made along the scanning direction of the
satellite, have typical individual precisions of 1--3\,mas each.
Therefore, they may contain valuable information on the orbital motion
that can be exploited to supplement the existing astrometry. We use
these measurements below, not only to improve the orbital elements but
also potentially to remove any biases in the original parallax and
proper motion determinations.  As in the case of the radial
velocities, the formal uncertainties for these \hip\ measurements,
and all other astrometric measurements described in this section, were
adjusted during our orbital analysis, as described next.

We note that \hstar\ is also being observed by the \gaia\ mission,
but the individual measurements are not expected to be publicly
available until the end of operations, several years from now.

\section{Orbital Analysis}
\label{sec:analysis}

Unlike all previous studies of \hstar AB, here we made use of all
observations simultaneously to solve for the orbital elements. This is
particularly helpful in this case because the astrometric information
is rather sparse, whereas the radial-velocity measurements are much
more numerous, they cover a larger number of orbital cycles, and they
therefore constrain the shape of the orbit and the ephemeris very
well. The astrometry's job is mostly to set the angular scale
and orientation of the orbit on the plane of the sky.

The elements of the relative orbit of \hstar AB are represented by the
standard elements $P$ (orbital period), $a^{\prime\prime}$ (angular
semimajor axis), $e$ (eccentricity), $i$ (inclination angle),
$\omega_{\rm B}$ (argument of periastron for the secondary), $\Omega$
(position angle of the ascending node), and $T$ (reference time of
periastron passage). The orbit of the photocentre is a scaled version
of the relative orbit, with angular semimajor axis $a_{\rm
  phot}^{\prime\prime}$.  As mentioned earlier, the secondary star is
so faint \citep[$\Delta V \approx 7.5$\,mag;][]{Henry:1993} that in
practice the photocentre coincides with the primary for observations
at optical wavelengths. Our orbital analysis in this section does not
require this assumption; we assume only that the location and motion
of the photocentre is the same for both the photographic measurements
of \cite{Lippincott:1983} and the \hip\ measurements (but see
below).  Two more elements were used to describe the spectroscopic
orbit: $K$ (velocity semiamplitude of the primary), and $\gamma$ (the
centre-of-mass velocity).

We have chosen to consider five additional adjustable parameters in
our analysis to account for possible differences in the
velocity zero points of the various spectroscopic data sets, relative
to one of them taken as the reference. The potential for these shifts
has often been overlooked in previous analyses of the spectroscopic
orbit of \hstar AB.  Given our efforts to carefully place the CfA velocities
(${\rm DS} + {\rm TRES}$) on the IAU system
(Section~\ref{sec:spectroscopy}), we chose those as the reference
data set. The five offsets ($\Delta RV_{\rm Camp}$, $\Delta RV_{\rm
  Beav}$, $\Delta RV_{\rm Halb}$, $\Delta RV_{\rm Abt}$, and $\Delta
RV_{\rm Fek}$,) correspond to the data sets of \cite{Campbell:1928},
\cite{Beavers:1986}, \cite{Halbwachs:2018}, \cite{Abt:2006}, and
\cite{Fekel:2018}, respectively, and are to be added to those velocities
to place them on the same system as the CfA measurements.

Our use of the \hip\ observations introduces another five
adjustable parameters that represent corrections to the position of
the barycentre ($\Delta\alpha^*$, $\Delta\delta$), the proper motion
components ($\Delta\mu_{\alpha}^*$, $\Delta\mu_{\delta}$), and the
parallax ($\Delta\pi_{\rm t}$) reported in the catalogue. The formalism
for incorporating the \hip\ intermediate data in an orbital fit follows the
description of \cite{Pourbaix:2000}, including the correlations
between measurements from the two independent data reduction consortia
\citep[see][]{ESA:1997}.

\setlength{\tabcolsep}{4pt}
\begin{deluxetable}{lc}
\tablewidth{0pc}
\tablecaption{Results of Our Orbital Analysis for \hstar AB\label{tab:results}}
\tablehead{
\colhead{~~~~~~~~~~~Parameter~~~~~~~~~~~} &
\colhead{Value}
}
\startdata
 $P$ (year)                              & $19.542 \pm 0.014$\phn    \\ 
 $T$ (yr)                                & $2016.702 \pm 0.012$\phn\phn\phn  \\ 
 $a^{\prime\prime}$ (arcsec)             & $0.6104 \pm 0.0097$     \\ 
 $a^{\prime\prime}_{\rm phot}$ (arcsec)  & $0.1329 \pm 0.0041$   \\ 
 $e$                                     & $0.4367 \pm 0.0020$   \\ 
 $\omega_{\rm B}$ (deg)                  & $27.15 \pm 0.35$\phn      \\ 
 $i$ (deg)                               & $100.36 \pm 0.89$\phn\phn     \\ 
 $\Omega$ (deg)                          & $32.25 \pm 0.85$\phn      \\ 
 $\gamma$ (\kms)                         & $+3.3672 \pm 0.0070$\phs  \\ 
 $K$ (\kms)                              & $2.7160 \pm 0.0072$   \\ [1ex]
 
 $\Delta RV_{\rm Camp}$ (\kms)           & $+0.61 \pm 0.34$\phs      \\ 
 $\Delta RV_{\rm Beav}$ (\kms)           & $-0.60 \pm 0.20$\phs      \\ 
 $\Delta RV_{\rm Halb}$ (\kms)           & $-0.024 \pm 0.049$\phs    \\ 
 $\Delta RV_{\rm Abt}$ (\kms)            & $+0.039 \pm 0.026$\phs    \\ 
 $\Delta RV_{\rm Fek}$ (\kms)            & $+0.071 \pm 0.013$\phs    \\ [1ex]

 $\Delta\alpha^*$ (mas)                  & $-21.1 \pm 1.9$\phn\phs       \\
 $\Delta\delta$ (mas)                  & $-82.4 \pm 2.7$\phn\phs       \\
 $\Delta\mu^*_{\alpha}$ (mas yr$^{-1}$)  & $+17.57 \pm 0.77$\phn\phs     \\ 
 $\Delta\mu_{\delta}$ (mas yr$^{-1}$)  & $+26.3 \pm 1.0$\phs\phn       \\ 
 $\Delta\pi_{\rm t}$ (mas)               & $-0.01 \pm 0.63$\phs      \\ [1ex]
\hline \\ [-1.5ex]
\multicolumn{2}{c}{Derived quantities} \\ [1ex]
\hline \\ [-1.5ex]
 $P$ (day)                               & $7137.6 \pm 5.1$\phm{222}      \\ 
 $T$ (HJD$-2,400,000$)                   & $57645.5 \pm 4.4$\phm{2222}     \\ [1ex]
 Total mass (M$_{\sun}$)                 & $1.204 \pm 0.066$     \\
 $q \equiv M_{\rm B}/M_{\rm A}$          & $0.268 \pm 0.013$     \\
 $M_{\rm A}$ (M$_{\sun}$)                & $0.95 \pm 0.11$       \\
 $M_{\rm B}$ (M$_{\sun}$)                & $0.254 \pm 0.019$     \\
 $a$ (au)                                & $7.72 \pm 0.14$       \\ [1ex]
 $\mu^*_{\alpha}$ (mas yr$^{-1}$)        & $+808.92 \pm 0.77$\phs\phn\phn    \\ 
 $\mu^*_{\delta}$ (mas yr$^{-1}$)        & $-153.8 \pm 1.0$\phs\phn\phn      \\ 
 $\pi_{\rm t}$ (mas)                     & $79.08 \pm 0.63$\phn      
\enddata
\end{deluxetable}
\setlength{\tabcolsep}{6pt}

We solved simultaneously for all orbital elements and auxiliary
parameters using standard non-linear least-squares techniques
\citep[e.g.,][]{Press:1992}.  The use of different types of
observations requires careful relative weighting for a balanced
solution. We handled this by applying multiplicative scaling factors
to the uncertainties, determined by iterations so as to achieve
reduced $\chi^2$ values near unity for each data set. For the WDS and
photographic observations this was done separately in each coordinate.
The results of our analysis may be found in Table~\ref{tab:results},
and the final multiplicative error scaling factors are given in
Table~\ref{tab:errors}. The bottom section of Table~\ref{tab:results}
lists various derived properties that we discuss in the next section.

\setlength{\tabcolsep}{10pt}
\begin{deluxetable}{lc}
\tablecaption{Scaling Factors for the Formal Uncertainties of the
  \hstar\ Measurements \label{tab:errors}}
\tablehead{
\colhead{~~~~~~~~~~~Data type~~~~~~~~~~~} &
\colhead{Factor}
}
\startdata
WDS position angles                &  1.24 \\
WDS separations                    &  0.99 \\
\cite{Lippincott:1983} $\Delta X$  &  1.09 \\
\cite{Lippincott:1983} $\Delta Y$  &  2.12 \\
\hip\ measurements                 &  0.98 \\
\cite{Campbell:1928} RVs           &  1.48 \\
\cite{Beavers:1986} RVs            &  0.96 \\
\cite{Halbwachs:2018} RVs          &  0.81 \\
\cite{Abt:2006} RVs                &  0.46 \\
\cite{Fekel:2018} RVs              &  1.31 \\
DS RVs                             &  2.23 \\
TRES RVs                           &  0.63
\enddata
\end{deluxetable}
\setlength{\tabcolsep}{6pt}

\begin{figure}
\epsscale{1.18}
\plotone{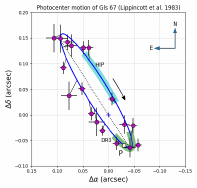}
\figcaption{Path of the photocentre on the plane of the sky. The
  points show the photographic observations by \cite{Lippincott:1983}
  with their final associated uncertainties. Line segments connect the
  observations with the predicted location on the orbit. The dotted
  line represents the line of nodes, and periastron is indicated with
  a square (`P'). The shaded regions correspond to the coverage of
  the \hip\ (`HIP') and \gaia\ (`DR3')
  missions. \label{fig:photorbit}}
\end{figure}

A graphical representation of the photographic observations by
\cite{Lippincott:1983} that trace the photocentre motion is presented in
Fig.~\ref{fig:photorbit}, with our best-fitted model. On the same plot
we indicate the part of the orbit in which \hip\ observations were
made, as well as the section covered by the measurements included in
the Third Data Release (DR3) of the \gaia\ mission
\citep{GaiaDR3:2022}, in which the star has the identifier 348515297330773120.

The \hip\ observations are seen to have been obtained on a side of
the orbit with little curvature, on which the motion of the primary
was directed toward the south and west relative to the barycentre.  The small curvature
suggests that any biases in the parameters reported in the \hip\
catalogue are more likely to be in the proper motion components 
than in the parallax. A disadvantage is that the small degree of
curvature also diminishes the power of these observations to constrain
the shape and scale of the orbit. As a test, we solved for separate
values of $a^{\prime\prime}_{\rm phot}$ for the \hip\ and
\cite{Lippincott:1983} measurements of the photocentre. We verified
that while the values are entirely consistent within the uncertainties
($0\farcs140 \pm 0\farcs030$ and $0\farcs1328 \pm 0\farcs0041$,
respectively), the \hip\ result is much poorer, and the
p.m.\ determinations are weakened as well. This motivated us to assume
a common value for $a^{\prime\prime}_{\rm phot}$, which has no impact
on any of the other orbital properties.

\begin{figure}
\epsscale{1.18}
\plotone{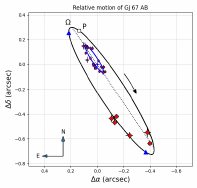}
\figcaption{Relative orbit of \hstar\ from our global model. Red
  symbols correspond to the WDS measurements that can be represented
  graphically, and the blue triangles mark the predicted locations for
  the 1992 and 2006 observations that lack a position angle. Short
  line segments connect the red dots with the corresponding calculated
  positions on the orbit.  The dotted line is the line of nodes (with
  the ascending node labelled $\Omega$), and periastron is indicated
  with a square labelled `P'. The motion of the photocentre is
  reproduced to scale from Fig.~\ref{fig:photorbit}, including the
  photographic measurements shown there.\label{fig:relorbit}.}
\end{figure}

The relative orbit of \hstar AB is presented in
Fig.~\ref{fig:relorbit} along with all complete WDS measurements, as well
as the N--S and E--W measurements for the 1982 epoch. The 1995 and
2006 observations lacking a position angle cannot be represented, but
their predicted locations are indicated with triangles. The
photocentre orbit is shown to scale along with the photographic
measurements from Fig.~\ref{fig:photorbit}.

A comparison of our orbital elements with all other determinations in the
literature is presented in Table~\ref{tab:orbits}.

\setlength{\tabcolsep}{5pt}
\begin{deluxetable*}{lccccccccc}
\tablewidth{0pc}
\tablecaption{Comparison of Published Orbital Solutions for \hstar AB \label{tab:orbits}}
\tablehead{
\colhead{Source} &
\colhead{$P$} &
\colhead{$a^{\prime\prime}$} &
\colhead{$e$} &
\colhead{$i$} &
\colhead{$\omega_{\rm A}$\tablenotemark{a}} &
\colhead{$\Omega$} &
\colhead{$T$\tablenotemark{b}} &
\colhead{$K$} &
\colhead{$\gamma$}
\\
\colhead{} &
\colhead{(yr)} &
\colhead{(\arcsec)} &
\colhead{} &
\colhead{(deg)} &
\colhead{(deg)} &
\colhead{(deg)} &
\colhead{(yr)} &
\colhead{(\kms)} &
\colhead{(\kms)}
}
\startdata
\cite{Lippincott:1983} & 19.50  &  \phm{\tablenotemark{c}}0.616\tablenotemark{c} &  0.42   &    104.0  &   100.0   &   32.6  &   2016.60    & \nodata  & \nodata  \\
                       & \phn0.28  &  0.040  &  0.06   &           &           &         &  \phn\phn\phn0.30    &          &          \\ [1ex]
                                                                                                                
\cite{Duquennoy:1991}  & 19.50  & \nodata &  0.42   &  \nodata  &   210.8   & \nodata &   2016.60    & 2.68     &  3.12    \\
                       & fixed  &         &  fixed  &           &  \phn\phn3.5   &         &   fixed      & 0.18     &  0.13    \\ [1ex]
                                                                                                                
\cite{Henry:1993}      & 19.50  &  0.565  &  0.42   &    104.0  &   200.0   &   32.6  &   2016.60    & \nodata  & \nodata  \\
                       & fixed  &  0.035  &  fixed  &    fixed  &   fixed   &  fixed  &   fixed      &          &          \\ [1ex]
                                                                                                                
\cite{Martin:1998}     & 19.50  &  0.565  &  0.42   &    104.0  &   100.0   &   32.6  &   2016.60    & \nodata  & \nodata  \\
                       & fixed  &         &  fixed  &    fixed  &   fixed   &  fixed  &   fixed      &          &          \\ [1ex]
                                                                                                                
\cite{Soderhjelm:1999} & 19.5   &  0.59   &  0.43   &    105    &   202     &   33    &   2016.6     & \nodata  & \nodata  \\ [1ex]
                                                                                                                
\cite{Abt:2006}        & 18.46  & \nodata &  0.34   &  \nodata  &   193     & \nodata &   2013.86    & 2.52     &  3.25    \\
                       &  \phn0.09  &         &  0.04   &           &  \phn\phn6     &         &  \phn\phn\phn0.34    & 0.16     &  0.10    \\ [1ex]
                                                                                                                
\cite{Miles:2017}\tablenotemark{d}   & 18.12  &  0.631  &  0.434  &     98.8  &   144.3   &  205.0  &   2011.75    & \nodata  & \nodata  \\ [1ex]
                                                                                                                
\cite{Fekel:2018}      & 19.550 & \nodata &  0.4474 &  \nodata  &   209.6   & \nodata &   2016.789   & 2.710    &  3.300   \\ 
                       &  \phn0.021 &         &  0.0051 &           &  \phn\phn1.0   &         &      \phn\phn\phn0.035   & 0.018    &  0.013   \\ [1ex]
                                                                                                                
This work              & 19.542 &  0.6104  &  0.4367 &    100.36 &   207.15  &   32.25 &   2016.702   & 2.7160   &  3.3672  \\
                       &  \phn0.014 &  0.0097  &  0.0020 &  \phn\phn0.89 &  \phn\phn0.35  &  \phn0.85 & \phn\phn\phn0.012   & 0.0072   &  0.0070 
\enddata

\tablenotetext{a}{This is the argument of periastron for the primary.
  The secondary values from \cite{Soderhjelm:1999} and our own have
  been shifted by 180\arcdeg, to facilitate the comparison.}

\tablenotetext{b}{Original epochs have been shifted by an integer
  number of periods to more closely match the reference time of
  periastron in this paper.}

\tablenotetext{c}{Value obtained by conversion of the linear semimajor
  axis given in the paper to angular measure, using the published
  parallax.}

\tablenotetext{d}{See comment at the end of Section~\ref{sec:relative}
  about a bias in this orbit. A 180\arcdeg\ shift in $\omega$ and
  $\Omega$ would bring the latter angle closer to other results, but
  $\omega$ would still be discrepant.}

\tablecomments{The second line for each entry contains the uncertainties.
  The correct value for the argument of periastron of
  \cite{Lippincott:1983} is $\omega_{\rm A} = 200\arcdeg$ \citep[see
    also][]{Henry:1992}, which reproduces the Thiele-Innes constants
  from her paper. Unfortunately, \cite{Martin:1998} adopted the
  erroneous value of 100\fdg0, and this biased their reanalysis of the
  \hip\ data giving unrealistically low masses for the components
  (Section~\ref{sec:introduction}), as they were also surprised to
  find.}

\end{deluxetable*}

\section{Discussion}
\label{sec:discussion}

Our mass determinations for the primary and secondary of \hstar AB are
consistent with those of \cite{Henry:1993}, and improve on the
uncertainties by at least a factor of two. As a check, we derived
another value for the mass by using stellar evolution models in conjunction
with our SPC determination of the spectroscopic properties in
Section~\ref{sec:spectroscopy}, which serves to also estimate the
radius of the star and other properties.

For this exercise we used the {\tt EXOFASTv2} code of
\cite{Eastman:2019}\footnote{\url{https://github.com/jdeast/EXOFASTv2}},
coupled with a fit to the spectral energy distribution (SED)
constrained by our parallax determination and brightness measurements
in the Johnson, Tycho-2, Sloan, 2MASS, and WISE systems
\citep{Mermilliod:1994, Hog:2000, Mallama:2014, Cutri:2003,
  Cutri:2012}. The light
contribution of the secondary was assumed to be insignificant for our
purposes.\footnote{This may not be quite true at the longer
  wavelengths. For example, \cite{Henry:1993}
  estimated a brightness difference in the $K$ band of about 
  4.4\,mag (see Table~\ref{tab:wds}). However, we verified that small corrections for this
  effect do not significantly change our results below.}  For the
model isochrones the {\tt EXOFASTv2} code relies on the MIST series of \cite{Choi:2016}, and
the spectroscopic quantities were used as priors in the fit. Fluxes
were based on the NextGen model atmospheres of \cite{Allard:2012}.
 
\begin{figure}
\epsscale{1.15}
\plotone{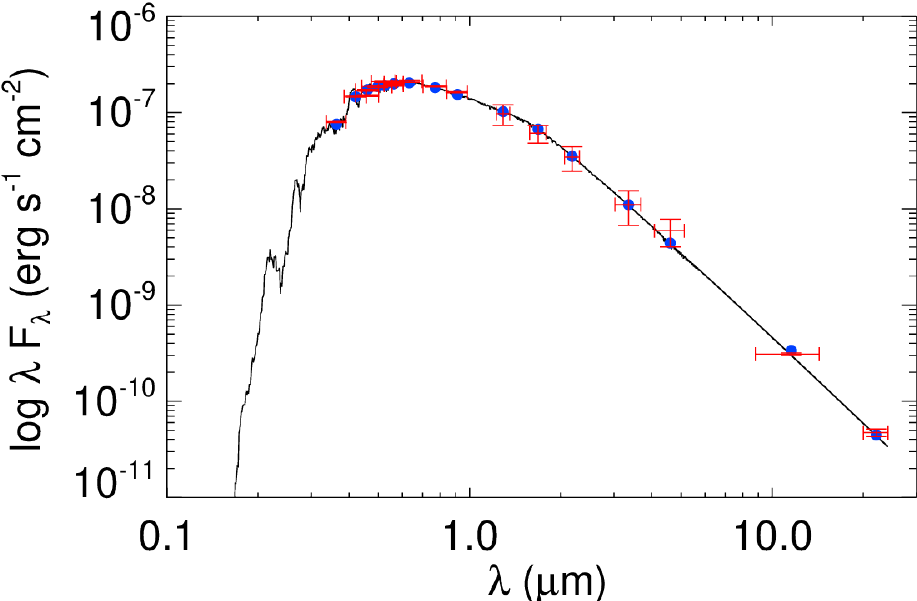}
\figcaption{Fit to the spectral energy distribution of \hstar\ with
  {\tt EXOFASTv2} \cite{Eastman:2019}. The curve is based on a model
  atmosphere from the NextGen series of \cite{Allard:2012} constrained
  by our SPC parameters from Section~\ref{sec:spectroscopy}. Red
  error bars represent the brightness measurements, and the blue dots correspond
  to the computed flux from the model.\label{fig:sed}}
\end{figure}

The resulting SED fit is shown in Fig.~\ref{fig:sed}, and the
primary mass we obtained is $M_{\rm A} = 0.986 \pm 0.069$\,M$_{\sun}$. This
is consistent with our formally less precise dynamical estimate from
the previous section. The radius of the primary is estimated to be $R_{\rm A} = 1.129
\pm 0.023$\,R$_{\sun}$, its luminosity is $L_{\rm A} = 1.353 \pm 0.034$\,L$_{\sun}$,
and the age according to the MIST models is $7.3 \pm 3.1$~Gyr.  The
main factor entering into the uncertainty in our dynamical mass
determination through Kepler's Third Law is the error in angular
semimajor axis of the orbit, which contributes about twice as much as
the error in the parallax. The period uncertainty has a negligible
contribution.

\begin{figure}
\epsscale{1.12}
\plotone{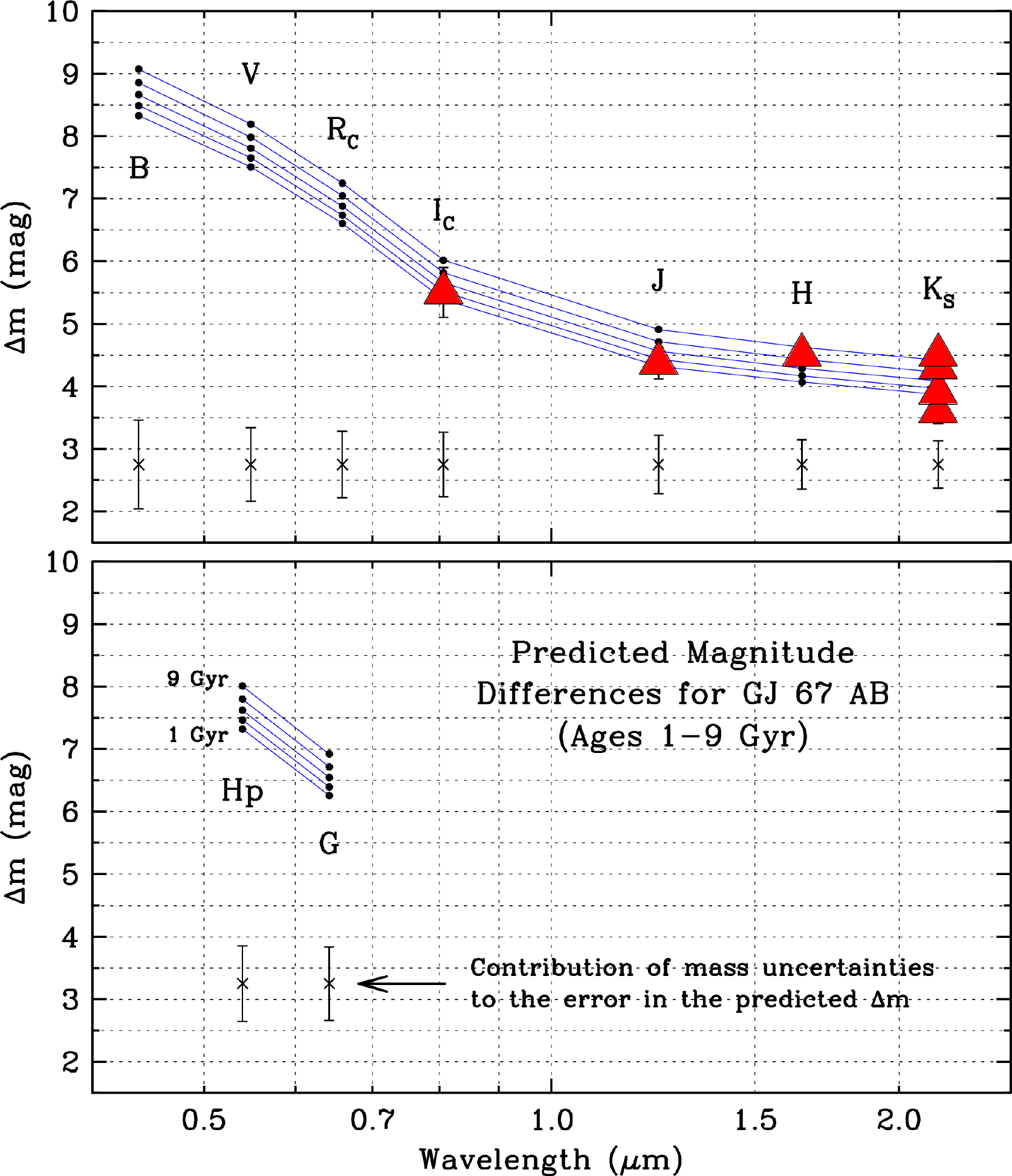}
\figcaption{{\it Top:} Predicted contrast between the primary and secondary
of \hstar AB compared with measurements from Table~\ref{tab:wds} (represented
with triangles), in magnitude units. Predictions in standard photometric bands
as labelled are based on solar-metallicity model isochrones from \cite{Chen:2014},
for ages of 1~Gyr (bottom) to 9~Gyr (top). Uncertainties in the predictions are
indicated along the bottom. {\it Bottom:} Same as above, for the
bandpasses of the space missions \hip\ (Hp) and \gaia\ (G).\label{fig:dmag}}
\end{figure}

As a further consistency check, we used our mass determinations from
Table~\ref{tab:results} along with stellar evolution models to compute
the brightness contrast between the primary and secondary in several standard
photometric filters, and compared those values with the measured magnitude differences
from Table~\ref{tab:wds}. For this we used model isochrones from the PARSEC
series by \cite{Chen:2014}, which have been shown by those authors to
perform better than others for low-mass stars such as the secondary of \hstar AB.
The results are presented in Figure~\ref{fig:dmag} for a range of ages
because the age of the system is not well constrained by the
observations, and the brightness of the primary changes by up to about 0.75~mag between
1 and 9~Gyr. The measured $\Delta m$ values in the top panel are seen
to be in good agreement with predictions. Uncertainties in the predictions
stemming from the mass errors are indicated schematically along the bottom of each panel.

The contrast at $V$ is roughly as estimated by \cite{Henry:1993}, while
the magnitude difference at $B$ (approximately the bandpass of the
photographic observations of \citealt{Lippincott:1983}) is larger, between 8 and 9~mag.
The light contribution of the secondary at these wavelengths is therefore
very small, so that for most practical purposes the measured semimajor axis
of the photocentre motion from the photographic observations can be considered to be equal to the
semimajor axis of the primary. A separate measure of the mass ratio
$M_{\rm B}/M_{\rm A}$ may then be derived as
$q_{\rm phot} \equiv a^{\prime\prime}_{\rm phot}/(a^{\prime\prime}-a^{\prime\prime}_{\rm phot}) =
0.278 \pm 0.012$. This estimate is independent of the parallax, but is
consistent with the value $q = 0.268 \pm 0.013$ listed in Table~\ref{tab:results} that 
relies on $\pi_{\rm t}$ through Kepler's Third Law and other properties. The predicted magnitude
differences for the space missions \hip\ and \gaia\ are given
in the lower panel of Figure~\ref{fig:dmag}. For \hip\ the secondary is
7--8~mag fainter than the primary; for \gaia\ it is about a magnitude brighter than that.

Concerning the parallax, we note that our result from the reanalysis
of the \hip\ intermediate data is not very different from the
values reported in both the original catalogue \citep{ESA:1997} and
the revision by \cite{vanLeeuwen:2007}. On the other hand, the
parallax entry in the \gaia\ DR3 catalogue \citep{GaiaDR3:2022} is very
different. We compare these values, along with others, in
Table~\ref{tab:pm}, which also lists proper motion determinations.

\setlength{\tabcolsep}{5pt}
\begin{deluxetable*}{lcccl}
\tablewidth{0pc}
\tablecaption{Proper Motion and Parallax Determinations for \hstar AB \label{tab:pm}}
\tablehead{
\colhead{Reference} &
\colhead{$\mu_{\alpha}^*$} &
\colhead{$\mu_{\delta}$} &
\colhead{$\pi_{\rm t}$} &
\colhead{Notes}
\\
\colhead{} &
\colhead{(mas yr$^{-1}$)} &
\colhead{(mas yr$^{-1}$)} &
\colhead{(mas)} &
\colhead{}
}
\startdata
\cite{Lippincott:1983}  &    \nodata                  &    \nodata       &  $72.1 \pm 2.8$\phn     & \\
\cite{Gliese:1995}      &    \nodata                  &    \nodata       &  $73.1 \pm 3.9$\phn     &  Third catalogue of nearby stars \\
\cite{vanAltena:1995}   &    \nodata                  &    \nodata       &  $74.2 \pm 4.4$\phn     &  Yale parallax catalogue, 4th ed. \\
\cite{ESA:1997}         &  $+791.35 \pm 0.65$\phn\phn\phs          &  $-180.16 \pm 0.47$\phs\phn\phn   &  $79.09 \pm 0.83$\phn    &  Original \hip\ catalogue \\
\cite{Martin:1998}      &    \nodata                  &    \nodata       &  $79.86 \pm 0.91$\phn   &  Based on \hip\ data \\
\cite{Soderhjelm:1999}  &    \nodata                  &    \nodata       &  $78.9 \pm 0.9$\phn     &  Based on \hip\ data \\
\cite{vanLeeuwen:2007}  &  $+791.47 \pm 0.48$\phn\phn\phs  &  $-180.80 \pm 0.36$\phs\phn\phn   &  $78.50 \pm 0.54$\phn    &  Revised \hip\ catalogue \\
\cite{GaiaDR2:2018}     &  $+813.34 \pm 0.63$\phn\phn\phs  &  $-171.03 \pm 0.76$\phs\phn\phn   &  $76.52 \pm 0.21$\phn    &  \gaia\ DR2 \\
\cite{GaiaDR3:2022}    &  $+824.76 \pm 0.41$\phn\phn\phs  &  $-156.42 \pm 0.39$\phs\phn\phn   &  $68.02 \pm 0.37$\phn    &  \gaia\ DR3 \\
\cite{Brandt:2021}      & $+805.869 \pm 0.026$\phn\phn\phs &  $-161.243 \pm 0.018$\phs\phn\phn &  \nodata                 & \gaia\ EDR3 + \hip \\ [1.5ex]
\cite{Hog:2000}         &  $+806.6 \pm 1.0$\phn\phn\phs    &  $-152.2 \pm 1.0$\phs\phn\phn     &   \nodata       &  Tycho-2 catalogue \\
\cite{Monet:2003}       &  +806\phs                        &  $-154$\phs           &   \nodata     &  USNO-B1.0 catalogue \\
\cite{Vondrak:2007}     &  $+806.60 \pm 0.77$\phn\phn\phs  &  $-154.17 \pm 0.38$\phs\phn\phn   &   \nodata       &  EOC-3 catalogue \\
\cite{Roser:2008}       &  $+807.2 \pm 1.6$\phn\phn\phs   &  $-156.3 \pm 1.8$\phs\phn\phn    &   \nodata       &  PPMX catalogue \\ [1.5ex]
This work               &  $+808.92 \pm 0.77$\phn\phn\phs  &  $-153.86 \pm 1.01$\phs\phn\phn   &  $79.08 \pm 0.63$\phn    &  Our reanalysis of \hip\ data
\enddata
\tablecomments{The uncertainties for the proper motions from \gaia\ DR2 and DR3 have been increased following \cite{Brandt:2018, Brandt:2021}. Note that the position, proper motions, and
parallax from the \gaia\ Early Third Data Release (EDR3) are identical to those in the final DR3.}
\end{deluxetable*}

The \gaia\ DR3 parallax is more than 10\,mas lower than ours, and lower
also than all other sources shown in the table. It is very different
as well from the value in the previous edition of the catalogue
\citep[\gaia\ DR2;][]{GaiaDR2:2018}. We believe the explanation has to
do with where in the orbit those observations were obtained. This is
shown by the shaded area in Fig.~\ref{fig:photorbit}. As opposed to
the situation with the \hip\ observations, which were gathered on
a part of the orbit with little curvature, the DR3 measurements
occurred precisely at the time when the photocentric path presents
the most curvature -- arguably the worst possible place for
determining the five standard astrometric parameters if the orbital
motion is not taken into account.\footnote{This bias could possibly
be alleviated by the addition of acceleration terms (proper motion derivatives) in the
\gaia\ solution, but \hstar\ is not one of the objects in the DR3
catalog for which this was attempted.}
Not surprisingly, the renormalised
unit weight error (RUWE) for the \gaia\ entry, which is a measure of
the quality of the
DR3 astrometric solution, is 2.892, much larger than typical
values for well-behaved astrometric solutions \citep[${\rm RUWE} \leq
  1.4$; see][]{Lindegren:2018}.  The DR3 results are therefore
suspect. Such a small value for the parallax ($68.02 \pm 0.37$\,mas),
together with our semimajor axis and period, would imply a primary
mass of about 1.48\,M$_{\sun}$, which is inconsistent with the spectral
type of the star.

Fig.~\ref{fig:photorbit} is also helpful to understand the
differences among the p.m.\ measurements in Table~\ref{tab:pm}. Both
editions of the \hip\ catalogue show a larger p.m.\ in declination
compared to ours, because the primary was moving southward at the
time, relative to the barycentre. The right ascension component is
smaller than ours, because the primary was moving toward the west.
During the \gaia\ observations, the net motion of the primary was toward
the east, and indeed the DR3 value of $\mu_{\alpha}^*$ is larger than
ours. On the other hand, the net motion in declination was probably
very small (southward for the first half, then northward), explaining
why the \gaia\ $\mu_{\delta}$ value is similar to ours. Most other
p.m.\ determinations in the table (Tycho-2, USNO-B1.0, EOC-3, PPMX),
which we have set aside from the others, happen to rely on positional
measurements that span decades, or up to a century in some cases,
which tends to average out the orbital motion. This is likely why they
more closely resemble our own estimates from the reanalysis of the \hip\ data,
which properly removed the effect.

As a final note, we draw attention to the proper motion determination
by \cite{Brandt:2021} in Table~\ref{tab:pm}, which is at least an order of
magnitude more precise than any of the others. This value was derived
from the positional difference between \hip\ and the \gaia\ Early Third
Date Release (EDR3)\footnote{The positions
from \gaia\ EDR3 are identical to those in DR3, as they are based on
the same observations.}, and can
potentially be used to constrain both the p.m.\ of the barycentre as well
as the orbital motion itself.
Here we have chosen not to use this information because we already make full
use of the \hip\ observations in our fit, as our main goal in
doing so was to rederive the trigonometric parallax free from the effects
of orbital motion. Nevertheless, as a sanity check it is of interest to
verify that the \hip-\gaia\ p.m.\ is consistent with our solution.
Dividing the change in the position of the photocentre at each epoch
by the epoch difference in each coordinate \citep[25.2~yr and 25.7~yr in R.A.\ and Dec,
respectively; see][]{Brandt:2021}
we obtain rates of change of $-2.52 \pm 0.22$ and $-6.01 \pm 0.39$ mas~yr$^{-1}$,
where the uncertainties account for the errors in all orbital elements involved,
and their correlations. Subtracting these values from the nominal
\hip-\gaia\ p.m.\ gives corrected values of $\mu_{\alpha,\rm H-G}^* =
+808.39 \pm 0.22$ and $\mu_{\delta, \rm H-G} = -155.23 \pm 0.39$~mas~yr$^{-1}$.
These differ from the motion derived from our reanalysis of the \hip\
measurements in Table~\ref{tab:pm} (bottom line) by just $0.53 \pm 0.80$ and
$1.37 \pm 1.08$~mas~yr$^{-1}$, indicating good agreement at the 0.7$\sigma$
and 1.3$\sigma$ significance levels.

\section{Concluding Remarks}
\label{sec:remarks}

\hstar\ is among the closest three dozen or so G dwarfs in the sky,
and as pointed out by \cite{Henry:1992}, half of them are binaries. While the
binary nature of \hstar\ has been known for nearly 50 yr, only partial orbital
solutions have been reported since then that have not made use of all
available measurements. In this paper we present our own spectroscopic
monitoring of the object spanning more than 35 yr, and combine it
with other velocities from the literature, and with all astrometric
observations of which we are aware. The latter include the reconstructed
photographic measurements by \cite{Lippincott:1983} that led to its
discovery as a binary, the few existing measures of the relative position,
and the more recent \hip\
intermediate data.  Our global fit to the observations leads to
improved mass estimates for both components, particularly for the M dwarf
secondary, and to what we expect to be a more accurate
determination of the parallax for the system, accounting for orbital
motion.

This work has provided evidence supporting the conclusion that the
parallax from the \gaia\ DR3 catalogue may be seriously biased. While the
individual \gaia\ measurements will not be available for reanalysis
until future data releases or until the end of the mission, by then the
coverage of the 19.5~yr orbit will be substantial, allowing for a much
improved determination not only of the parallax, but of other orbital
elements of the photocentric orbit as well. To complement those
observations, it is hoped that observers will also continue monitoring
the system with spatially resolved observations -- challenging as they are --
to better constrain the scale of the relative orbit, which is the most
poorly determined of its properties.

\begin{acknowledgements}

The spectroscopic observations of \hstar AB at the CfA reported were
obtained with the assistance of M.\ Calkins, J.\ Caruso, P.\ Berlind,
R.\ J.\ Davis, G.\ Esquerdo, D.\ W.\ Latham, R.\ P.\ Stefanik, and
J.\ Zajac. We thank them all. We are also grateful to R.\ J.\ Davis
and J.\ Mink for maintaining the echelle databases at the CfA, and to
R.\ Matson for providing a listing of the measurements of \hstar\ from
the Washington Double Star Catalogue (USNO). The referee, A.\
Tokovinin, is also thanked for helpful comments that improved the manuscript.
This research has made use
of the SIMBAD and VizieR databases, operated at the CDS, Strasbourg,
France, of NASA's Astrophysics Data System Abstract Service, and of
the Washington Double Star Catalogue maintained at the U.S.\ Naval
Observatory. The work has also made use of data from the European
Space Agency (ESA) mission \gaia\
(\url{https://www.cosmos.esa.int/gaia}), processed by the \gaia\ Data
Processing and Analysis Consortium (DPAC,
\url{https://www.cosmos.esa.int/web/gaia/dpac/consortium}). Funding
for the DPAC has been provided by national institutions, in particular
the institutions participating in the \gaia\ Multilateral Agreement.
This publication also made use of data products from the Wide-field
Infrared Survey Explorer, which is a joint project of the University
of California, Los Angeles, and the Jet Propulsion
Laboratory/California Institute of Technology, funded by the National
Aeronautics and Space Administration.

\end{acknowledgements}

\section{Data Availability}

The data underlying this article are available in the article and in its online supplementary material.


\begin{thebibliography}

\bibitem[Abt \& Willmarth(2006)]{Abt:2006} Abt, H.~A. \& Willmarth,
  D.\ 2006, \apjs, 162, 207

\bibitem[Allard et al.(2012)]{Allard:2012} Allard, F., Homeier, D., \&
  Freytag, B.\ 2012, Philosophical Transactions of the Royal Society
  of London Series A, 370, 2765

\bibitem[Allende Prieto et al.(2004)]{Allende:2004} Allende Prieto,
  C., Barklem, P.~S., Lambert, D.~L., et al.\ 2004, \aap, 420, 183

\bibitem[Beavers \& Eitter(1986)]{Beavers:1986} Beavers, W.~I. \&
  Eitter, J.~J.\ 1986, \apjs, 62, 147.

\bibitem[Boeche \& Grebel(2016)]{Boeche:2016} Boeche, C. \& Grebel,
  E.~K.\ 2016, \aap, 587, A2
  
\bibitem[Brandt(2018)]{Brandt:2018} Brandt, T.~D.\ 2018, \apjs, 239, 31

\bibitem[Brandt(2021)]{Brandt:2021} Brandt, T.~D.\ 2021, \apjs, 254, 42

\bibitem[Buchhave et al.(2012)]{Buchhave:2012} Buchhave, L.~A.,
  Latham, D.~W., Johansen, A., et al.\ 2012, \nat, 486, 375

\bibitem[Campbell(1928)]{Campbell:1928} Campbell, W.~W.\ 1928,
  Publications of Lick Observatory, 16, 1
  
\bibitem[Chen et al.(2014)]{Chen:2014} Chen, Y., Girardi, L., Bressan,
  A., et al.\ 2014, \mnras, 444, 2525

\bibitem[Choi et al.(2016)]{Choi:2016} Choi, J., Dotter, A., Conroy,
  C., et al.\ 2016, \apj, 823, 102

\bibitem[Cutri et al.(2003)]{Cutri:2003} Cutri, R.~M., Skrutskie,
  M.~F., van Dyk, S., et al.\ 2003, The IRSA 2MASS All-Sky Point
  Source Catalog, NASA/IPAC Infrared Science Archive,
  \url{http://irsa.ipac.caltech.edu/applications/Gator/}

\bibitem[Cutri et al.(2012)]{Cutri:2012} Cutri, R.~M. \& et
  al.\ 2012, WISE All-Sky Data Release, VizieR Online Data Catalog,
  II/311

\bibitem[Duquennoy \& Mayor(1991)]{Duquennoy:1991} Duquennoy, A. \&
  Mayor, M.\ 1991, \aap, 248, 485

\bibitem[Eastman et al.(2019)]{Eastman:2019} Eastman, J.~D.,
  Rodriguez, J.~E., Agol, E., et al.\ 2019, arXiv:1907.09480

\bibitem[ESA(1997)]{ESA:1997} ESA, ed.\ 1997, ESA Special Publication,
  Vol.\ 1200, The \hip\ and Tycho Catalogues

\bibitem[Fekel et al.(2018)]{Fekel:2018} Fekel, F.~C., Willmarth,
  D.~W., Abt, H.~A., et al.\ 2018, \aj, 156, 117

\bibitem[F\H{u}r\'esz(2008)]{Furesz:2008} F\H{u}r\'esz, G. 2008, PhD
  thesis, Univ.\ Szeged, Hungary

\bibitem[\gaia\ Collaboration et al.(2018)]{GaiaDR2:2018} \gaia\
  Collaboration, Brown, A.~G.~A., Vallenari, A., et al.\ 2018, \aap,
  616, A1


\bibitem[\gaia\ Collaboration et al.(2022)]{GaiaDR3:2022} \gaia\
Collaboration, Vallenari, A., et al.\ 2022, \aap, in press

\bibitem[Gliese \& Jahreiss(1995)]{Gliese:1995} Gliese, W. \&
  Jahreiss, H.\ 1995, VizieR Online Data Catalog, V/70A

\bibitem[Halbwachs et al.(2018)]{Halbwachs:2018} Halbwachs, J.-L., Mayor, M., \& Udry, S.\ 2018, \aap, 619, A81

\bibitem[Hartkopf et al.(1997)]{Hartkopf:1997} Hartkopf, W.~I.,
  McAlister, H.~A., Mason, B.~D., et al.\ 1997, \aj, 114, 1639

\bibitem[Henry \& McCarthy(1993)]{Henry:1993} Henry, T.~J. \&
  McCarthy, D.~W.\ 1993, \aj, 106, 773

\bibitem[Henry et al.(1992)]{Henry:1992} Henry, T.~J., McCarthy,
  D.~W., Freeman, J., et al.\ 1992, \aj, 103, 1369

\bibitem[H{\o}g et al.(2000)]{Hog:2000} H{\o}g, E., Fabricius, C.,
  Makarov, V.~V., et al.\ 2000, \aap, 355, L27

\bibitem[Latham(1992)]{Latham:1992} Latham, D.\ W. 1992, in IAU
  Coll.\ 135, Complementary Approaches to Double and Multiple Star
  Research, ASP Conf.\ Ser.\ 32, eds.\ H.\ A.\ McAlister \&
  W.\ I.\ Hartkopf (San Francisco: ASP), 110

\bibitem[Latham et al.(2002)]{Latham:2002} Latham, D.\ W., Stefanik,
  R.\ P., Torres, G., et al.\ 2002, \aj, 124, 1144

\bibitem[Lindegren(2018)]{Lindegren:2018} Lindegren L., 2018,
  Re-normalising the astrometric chi-square in \gaia\ DR2,
  GAIA-C3-TN-LU-LL-124,
  \url{https://dms.cosmos.esa.int/COSMOS/doc\_fetch.php?id=3757412}

\bibitem[Lippincott \& Lanning(1976)]{Lippincott:1976} Lippincott,
  S.~L. \& Lanning, J.~J.\ 1976, \baas, 8, 360

\bibitem[Lippincott et al.(1983)]{Lippincott:1983} Lippincott, S.~L.,
  Braun, D., \& McCarthy, D.~W.\ 1983, \pasp, 95, 271

\bibitem[Mallama(2014)]{Mallama:2014} Mallama, A.\ 2014, \jaavso, 42,
  443

\bibitem[Martin et al.(1998)]{Martin:1998} Martin, C., Mignard, F., Hartkopf, W.~I., et al.\ 1998, \aaps, 133, 149

\bibitem[Mason et al.(2001)]{Mason:2001} Mason, B.~D., Wycoff, G.~L.,
  Hartkopf, W.~I., et al.\ 2001, \aj, 122, 3466

\bibitem[Mermilliod(1994)]{Mermilliod:1994} Mermilliod, J.-C.\ 1994,
  Bulletin d'Information du Centre de Donnees Stellaires, 45, 3

\bibitem[Miles \& Mason(2017)]{Miles:2017} Miles, S.\ K.\ N., \&
  Mason, B.\ D. 2017, in IAU Double Star Inf.] Circ., 191, 1

\bibitem[Monet et al.(2003)]{Monet:2003} Monet, D.~G., Levine, S.~E.,
  Canzian, B., et al.\ 2003, \aj, 125, 984

\bibitem[Nordstr\"om et al.(1994)]{Nordstrom:1994} Nordstr\"om, B.,
  Latham, D.\ W., Morse, J.\ A., et al.\ 1994, \aap, 287, 338

\bibitem[Pourbaix \& Jorissen(2000)]{Pourbaix:2000} Pourbaix, D. \&
  Jorissen, A.\ 2000, \aaps, 145, 161

\bibitem[Press et al.(1992)]{Press:1992} Press, W.\ H., Teukolsky,
  S.\ A., Vetterling, W.\ T., \& Flannery, B.\ P. 1992, Numerical
  Recipes, (2nd.\ Ed.; Cambridge: Cambridge Univ.\ Press), 650

\bibitem[Ram{\'\i}rez et al.(2007)]{Ramirez:2007} Ram{\'\i}rez, I.,
  Allende Prieto, C., \& Lambert, D.~L.\ 2007, \aap, 465, 271

\bibitem[Roberts(2011)]{Roberts:2011} Roberts, L.~C.\ 2011, \mnras,
  413, 1200

\bibitem[R{\"o}ser et al.(2008)]{Roser:2008} R{\"o}ser, S., Schilbach,
  E., Schwan, H., et al.\ 2008, \aap, 488, 401

\bibitem[Serabyn et al.(2007)]{Serabyn:2007} Serabyn, E., Wallace, K.,
  Troy, M., et al.\ 2007, \apj, 658, 1386

\bibitem[S{\"o}derhjelm(1999)]{Soderhjelm:1999} S{\"o}derhjelm, S.\ 1999, \aap, 341, 121

\bibitem[Stefanik et al.(1999)]{Stefanik:1999} Stefanik, R.\ P.,
  Latham, D.\ W., \& Torres, G.\ 1999, in ASP Conf.\ Ser.\ 185, IAU
  Coll.\ 170, Precise Stellar Radial Velocities,
  ed.\ J.\ B.\ Hearnshaw \& C.\ D.\ Scarfe (San Francisco, CA: ASP)
  354

\bibitem[Szentgyorgyi \& F\H{u}r\'esz(2007)]{Szentgyorgyi:2007}
  Szentgyorgyi, A.\ H., \& F\H{u}r\'esz, G. 2007, RMxAC, 28, 129

\bibitem[van Altena et al.(1995)]{vanAltena:1995} van Altena, W.~F.,
  Lee, J.~T., \& Hoffleit, E.~D.\ 1995, New Haven, CT: Yale University
  Observatory, C1995, 4th ed., completely revised and enlarged

\bibitem[van Leeuwen(2007)]{vanLeeuwen:2007} van Leeuwen, F. 2007,
  Astrophysics Space Science Library, Vol. 350, \hip, the New
  Reduction of the Raw Data (Berlin: Springer)

\bibitem[Vondr{\'a}k \& {\v{S}}tefka(2007)]{Vondrak:2007} Vondr{\'a}k,
  J. \& {\v{S}}tefka, V.\ 2007, \aap, 463, 783

\bibitem[Worley \& Douglass(1997)]{Worley:1997} Worley, C.~E. \&
  Douglass, G.~G.\ 1997, \aaps, 125, 523

\end{thebibliography}
\end{document}